\documentstyle[aps,epsf,prl]{revtex}
\begin{document}

\title{Spectroscopic fingerprints of a surface Mott-Hubbard insulator: 
the case of SiC(0001)}
\author{G. Santoro$^{1,2}$, S. Scandolo$^{1,2}$, E. Tosatti$^{1,2,3}$}
\address{
$^1$International School for Advanced Studies (SISSA), Via Beirut 2,
Trieste, Italy\\
$^2$Istituto Nazionale per la Fisica della Materia (INFM), Via Beirut 2, 
Trieste, Italy\\
$^3$International Centre for Theoretical Physics (ICTP), Trieste, Italy}
\maketitle

\begin{abstract}
We discuss the spectroscopic fingerprints that a surface Mott-Hubbard
insulator should show at the intra-atomic level. The test case considered
is that of the Si-terminated SiC(0001)$\sqrt{3}\times\sqrt{3}$ surface, 
which is known experimentally to be insulating.
We argue that, due to the Mott-Hubbard phenomenon, spin unpaired electrons 
in the Si adatom dangling bonds are expected to give
rise to a Si-2p core level spectrum with a characteristic three-peaked
structure, as seen experimentally. This structure results from the joint 
effect of intra-atomic exchange, spatial anisotropy, and spin-orbit coupling. 
Auger intensities are also discussed. 
\end{abstract}

\vspace{3mm}

The fractional adlayer structures on semiconductors show rich 
phase diagrams with potential instabilities of charge density 
wave (CDW) and spin density wave (SDW) type 
at low temperatures, as well as Mott insulating phases \cite{santoro}.
The (0001)$\sqrt{3}\times \sqrt{3}$ surface of hexagonal SiC,
(as well as the closely analogous (111)$\sqrt{3}\times \sqrt{3}$ of 
cubic SiC) expected from standard LDA calculations to be a 2D metal with 
a half-filled narrow band of surface states in the bulk energy gap
\cite{SiC-struc,pollmann} --
is instead experimentally proven to be an insulator with a rather large 
($2$ eV) band gap \cite{johansson,IPES}.
It has been suggested that this system is a Mott 
insulator\cite{neugebauer} due to the large value of ratio $U/W$, 
where $U$ is the Coulomb interaction parameter, of the order of several eV, 
and $W$ is the surface bandwidth, calculated to be about $0.35$ eV
\cite{SiC-struc,pollmann}. 
Very recent STM data further confirm this picture \cite{rama}. 

A recent LSDA+U calculation\cite{anisimov_sic} predicts 
an antiferromagnetic insulating state with $3\times 3$ magnetic 
periodicity and a indirect gap 
of about $1.5$eV, in reasonable agreement with the experimental 
value \cite{johansson,IPES}. 
An exchange coupling $J$ between neighboring Si adatoms of about $30K$ 
is also calculated. 
The LSDA+U magnetic state, with spin-collinearity enforced in the calculation, 
turns out to have a uniform magnetization $m^z=1/3$: 
one spin-down and two spin-up adatom surface bands are completely filled 
in the $3\times 3$ Brillouin zone. 
Crudely speaking, this corresponds to two adatom dangling-bond orbitals having
spin-up electrons and one having spin-down electron in 
each triangular plaquette. 
When spin non-collinearity is allowed one finds, {\em e.g.,}
from Hartree-Fock Hubbard model
calculations,\cite{anisimov_sic} that the state with spins lying on a plane 
-- in a $120^o$ three-sublattice N\'eel structure, typical of the triangular 
lattice Heisenberg antiferromagnet\cite{noncol} -- has slightly better energy than the 
collinear one, but the resulting energy gain is so small, 
about $0.5$ meV/adatom, that unrealistically precise ab-initio calculations 
would be necessary to decide which of the two is the actual ground state. 
Moreover, the bands for the two solutions are rather similar,
apart from some extra splittings introduced by the collinear magnetic solution
\cite{anisimov_sic}.

In reality, both the LSDA+U and the Hartree-Fock Hubbard calculations, 
being mean-field approaches are at best only cartoons for the actual 
Mott-Hubbard state, which is insulating not because it is magnetic, 
but rather due to strong correlations inhibiting electron hopping. 
Once this is appreciated, one realizes that -- even in the absence of 
any true magnetic long-range order -- finite magnetic moments on the 
Si adatoms should be experimentally detectable up to high temperatures, 
so long as there is an insulating gap. 
The low-temperature low-energy collective physics of these unpaired adatom
spins is instead governed by a much smaller energy scale, 
the exchange coupling $J$.
Is it conceivable that a genuine low-temperature spin long-range order (LRO) 
would be detectable in this quasi-2D system? 
It would seem unlikely, since:
1) The estimates for $J$ already suggest that LRO, if present, must be there
only at very low temperature. Moreover, thermal fluctuations might destroy LRO
altogether, if the continuous spin symmetry is not broken by anisotropy
effects due to spin-orbit coupling;
2) It is at present unclear which spin ordering will be stabilized by the 
spin-orbit coupling; 
3) If spins lie on a plane, with zero net magnetization, they are presumably 
hard to detect with LRO sensitive techniques.
Therefore, without ruling out totally the possibility of 
low-temperature LRO, we 
believe that the SiC(0001)$\sqrt{3}\times \sqrt{3}$ 
surface will most likely be in an overall paramagnetic Mott insulating state,
in spite of the existence of an on-site moment, at least down to liquid 
nitrogen temperature.

It is the aim of the present paper to discuss in some detail what kind of 
experimental signatures one should look for in a surface 
Mott-Hubbard insulator, apart from a gap in the photoemission spectrum.
As mentioned previously, methods which require magnetic LRO are likely 
not viable. 
The situation could be much more promising at the intra-atomic level. 
We will argue below that the very fact that the system is in a Mott-Hubbard
insulating state, independently of the magnetic LRO issue, brings about 
unpaired spins in the adatom dangling bonds, which in turn determine 
exchange splittings in the adatom core levels which should be large and 
detectable. 
These splittings may in fact have been already observed for SiC(0001) in the 
surface component of the Si-2p core level photoemission spectrum,\cite{core}
although attributed so far to some unspecified charge inequivalence 
of two types of Si adatoms. 

To make things more concrete, let us assume that, due to the Mott-Hubbard 
phenomenon, the $3p_z$-like dangling bond orbital of the Si adatom in the
$\sqrt{3}\times \sqrt{3}$ SiC(0001) is singly occupied by an unpaired
electron, say of spin up. 
In reality, the Wannier function for the surface state is not a pure $3p_z$, 
since it extends itself down into the first SiC 
bilayer\cite{SiC-struc,anisimov_sic}: 
the weight $\alpha$ of the Si $3p_z$ orbital in the Wannier function is about 
$0.5$, the remaining lobe having a negligeable overlap with the Si-adatom 
core \cite{SiC-struc,anisimov_sic}. 
The $3s$ and $3p_{x,y}$ orbitals of the Si-adatom are involved in the
bonding with first-layer C atoms, and contain spin-paired electrons. 
Therefore, a Si-adatom has two distinct sources of anisotropy:
i) the first comes about because the occupation $\alpha$ 
of the $3p_z$ orbital is not equal to the occupation, call it $\beta$, of the
$3p_x$ or $3p_y$ orbitals; 
ii) the second, and most important, source of anisotropy
is due to the fact that the electron in the $3p_z$ dangling bond is
{\em spin unpaired\/}, as opposed to the standard singlet bonding configuration
in $3p_{x,y}$. 
These anisotropies break the core level symmetry between $2p_z$ 
and $2p_{x,y}$ by an amount
which we will argue to be not negligeable compared to spin-orbit.
We will now estimate the expected $2p$ core level splittings, with and
without a spin-orbit coupling, and show that while spatial anisotropy of type
i) by itself would result in a very small effect, the spin 
induced anisotropy ii) has a large influence on the spectrum. 

\begin{figure}
\centerline{\epsfxsize=10cm %
\epsfbox{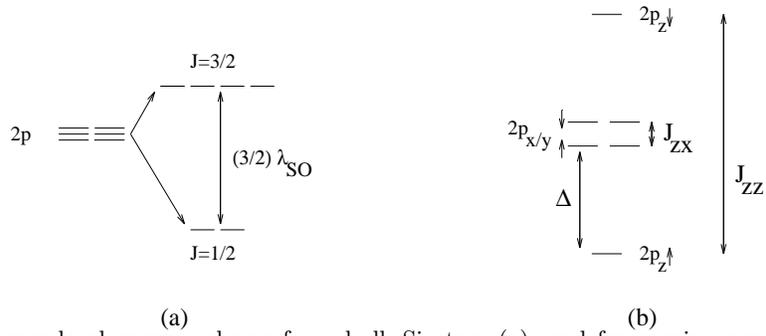}
} 
\caption{
Schematic 2p core level energy scheme for a bulk Si atom (a), and for a 
spin-unpaired surface Si adatom (b). Estimates for the splittings in
(b), discussed in the text, are $\Delta\approx 0.5$eV, $J_{zz}\approx 1$eV,
and $J_{zx}\approx 0.1$eV. 
}
\label{spectrum:fig}
\end{figure}

\begin{figure}
\centerline{\epsfxsize=12cm%
\epsfbox{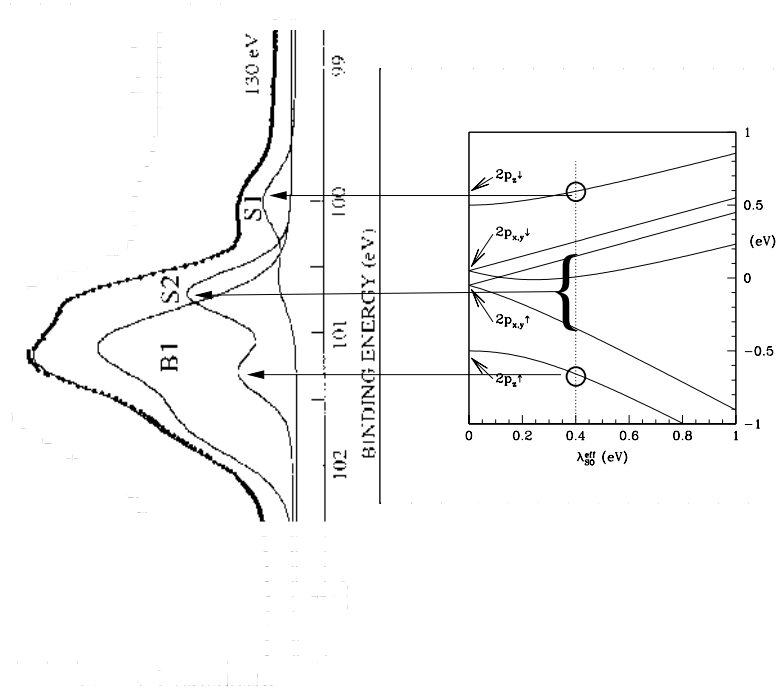}
} 
\caption{
Si 2p core levels calculated as a function of spin-orbit coupling,
in comparison with photoemission data and original fitting in terms
of bulk (B1) and two surface sites S1 and S2 by Johansson 
\protect{\it et al.\/}\protect\cite{core}. 
Upon inclusion of intra-atomic exchange splitting the 
whole surface contribution S1+S2 can be explained as due to a single site. 
This figure is taken from Ref.\ \protect\cite{anisimov_sic}. 
}
\label{levels:fig}
\end{figure}

The calculation we perform is very simple, and is based on the 
Si $2p-3p$ Slater exchange and Coulomb integrals, obtained from pure
atomic Hartree-Fock calculations. 
Assume the Si adatom valence electrons to be distributed with weights 
$\alpha$ in $3p_z\uparrow$, and $\beta/2$ in each $3p_{x,y}\uparrow$ and $\downarrow$. 
Let us consider first the situation in absence of spin-orbit coupling, 
$\lambda_{\rm SO}=0$. 
The Hartree-Fock energy of the $2p_z\uparrow$ states can be written as
\begin{equation} \label{2pz:eqn}
\epsilon_{2p_z\uparrow} = \alpha [K_{2p_z,3p_z}-J_{2p_z,3p_z}] 
+ 2 \frac{\beta}{2} [ 2 K_{2p_z,3p_x}-J_{2p_z,3p_x}] + \dots \;,
\end{equation}
where the dots stand for one-body terms and two-body contributions due to 
other spin-paired occupied orbitals which will exactly cancel out in all 
the energy differences which we will calculate. Here 
$K_{2p_z,3p_z(x)} = 
\left( 2p_z, 3p_{z(x)} | e^2/|{\bf r}_1-{\bf r}_2| | 2p_z, 3p_{z(x)} \right)$, 
and 
$J_{2p_z,3p_z(x)} = 
\left( 2p_z, 3p_{z(x)} | e^2/|{\bf r}_1-{\bf r}_2| | 3p_{z(x)}, 2p_z \right)$, 
are the relevant $2p-3p$ Coulomb and exchange Slater integrals. 
The factor $2$ in front of the $\beta/2$ term is due to equal contributions
coming from $3p_y$ and $3p_x$. 
The factor $2$ in front of $K_{2p_z,3p_x}$ keeps into account the
direct contribution due to both $\uparrow$ and $\downarrow$ $3p_{x,y}$ 
electrons. (The corresponding exchange term has no factor $2$ because only
$3p_{x,y}$ $\uparrow$ electron contribute to the exchange gain of the
$2p_z\uparrow$ level.)
A similar calculation for $\epsilon_{2p_x\uparrow}=\epsilon_{2p_y\uparrow}$
gives  
\begin{equation} \label{2px:eqn}
\epsilon_{2p_x\uparrow} = \alpha [K_{2p_x,3p_z}-J_{2p_x,3p_z}] 
+ \frac{\beta}{2} [ 2 K_{2p_x,3p_x}-J_{2p_x,3p_x}] 
+ \frac{\beta}{2} [ 2 K_{2p_x,3p_y}-J_{2p_x,3p_y}] 
+ \dots \;,
\end{equation}
which can be reexpressed, due to rotational symmetry, in terms of the
previously introduced $K_{2p_z,3p_z(x)}$ and $J_{2p_z,3p_z(x)}$. 
We now estimate the Slater integrals assuming pure atomic
hydrogen-like $2p-3p$ orbitals 
(the effect of intra-atomic hybridization in a full
atomic HF calculation would change a bit the numbers: these effects
will be estimated later on). 
Numerically, the two different Coulomb integrals appearing in 
Eqs.\ \ref{2pz:eqn}-\ref{2px:eqn} are large and comparable
($K_{2p_z,3p_z} \approx 30.5$ eV, $K_{2p_z,3p_x} \approx 28.2$ eV),
whereas the $2p_z-3p_z$ exchange integral is smaller
($J_{2p_z,3p_z} \approx 4.3$ eV), but one order of magnitude larger than
the $2p_z-3p_x$ exchange integral 
($J_{2p_z,3p_x} \approx 0.5$ eV). 
The large difference between $J_{2p_z,3p_z}$ and $J_{2p_z,3p_x}$ 
immediately implies that the exchange splitting between $2p_z\uparrow$
and $2p_z\downarrow$ is rather large
\[ \epsilon_{2p_z\downarrow} - \epsilon_{2p_z\uparrow} \approx \alpha
J_{2p_z,3p_z} \approx 2 eV \] 
whereas the correponding exchange splitting between $2p_{x/y}\uparrow$
and $2p_{x/y}\downarrow$ is one order of magnitude smaller:
\[ \epsilon_{2p_{x/y}\downarrow} - \epsilon_{2p_{x/y}\uparrow} 
\approx \alpha J_{2p_z,3p_x} \approx 0.2 eV \] 
We can also calculate, using Eqs.\ \ref{2pz:eqn}-\ref{2px:eqn}, 
the splitting 
$\Delta=\epsilon_{2p_{x/y}\uparrow}-\epsilon_{2p_z\uparrow}$
between the $2p_z\uparrow$ and the $2p_{x/y}\uparrow$ 
levels, 
\begin{equation} \label{Delta:eqn}
\Delta = \alpha [(J_{2p_z,3p_z}-J_{2p_z,3p_x}) -
           (K_{2p_z,3p_z}-K_{2p_z,3p_x}) ] + 
   \frac{\beta}{2} [ 2(K_{2p_z,3p_z}-K_{2p_z,3p_x}) -
           (J_{2p_z,3p_z}-J_{2p_z,3p_x}) ]  \;.
\end{equation}
Notice that the structure of the $\alpha$ contribution in Eq.\ \ref{Delta:eqn}
differs from that of the $\beta$ term because the $3p_z$ is assumed to
be spin unpaired (anisotropy ii), above): If we were to assume that the
occupancy $\alpha$ was due $\alpha/2$ $3p_z\uparrow$ and 
$\alpha/2$ $3p_z\downarrow$, the $\alpha$ term would be given by
$(\alpha/2) [(J_{2p_z,3p_z}-J_{2p_z,3p_x})-2(K_{2p_z,3p_z}-K_{2p_z,3p_x})]$ 
and would exactly cancel off the $\beta$ term when $\alpha=\beta$ (bulk case). 
Plugging in the numbers for the Coulomb and exchange integrals we
get $[2(K_{2p_z,3p_z}-K_{2p_z,3p_x})-(J_{2p_z,3p_z}-J_{2p_z,3p_x})] \approx
[2(2.3 eV)-3.8 eV]=0.6 eV$, and 
$[(J_{2p_z,3p_z}-J_{2p_z,3p_x})-(K_{2p_z,3p_z}-K_{2p_z,3p_x})] \approx
[3.8 eV - 2.3 eV]=1.5 eV$,  which provide the desidered estimate for
$\Delta \approx \alpha [ 1.5 eV ] + (\beta/2)[0.6 eV]$, 
i.e., $\Delta\approx 1 eV$ for $\alpha\approx 0.5$ and $\beta\approx 1$. 
The effect of intra-atomic orbital hybridization which modifies the
atomic Hartree-Fock orbitals with respect to pure hydrogen-like $2p-3p$ 
atomic orbitals, is estimated, from a full Hartree-Fock $Si^+$ atomic 
calculation, to further reduce $\Delta$ and the $2p_{z/x}$ exchange splittings
by a factor of about $2$, so that we arrive at the crude estimate 
$\Delta\approx 0.5 eV$,
$\epsilon_{2p_z\downarrow}-\epsilon_{2p_z\uparrow}\approx 1 eV$, and
$\epsilon_{2p_x\downarrow}-\epsilon_{2p_x\uparrow}\approx 0.1 eV$. 
The overall picture for the Si-adatom $2p$ spectrum in absence of spin-orbit,
but with all the previously described anisotropy effects included, is
summarized in Fig.\ \ref{spectrum:fig}(b).
When we include the spin-orbit interaction 
$\lambda_{\rm SO} {\bf L} \cdot {\bf S}$, the final core hole levels
are obtained as in Fig.\ \ref{levels:fig}. 
For $\lambda_{\rm SO}=0.4$ eV 
(the value extracted from the bulk $2p_{J=1/2}-2p_{J=3/2}$ experimental
core level splitting) we find a roughly three-peaked structure,
the broad central peak four times as strong as each side peak.
This offers an alternative explanation of the experimental lineshape\cite{core}
(see Fig.\ \ref{levels:fig}) in terms of a single exchange-split multiplet,
rather than two chemically inequivalent sites $S_1$ and $S_2$ \cite{core} 
whose existence is otherwise not supported. 
Incidentally, a similar three-peaked structure for the Si-2p core level
spectrum has been experimentally observed for a sub-monolayer Si segregated
at the surface of iron\cite{rossi}. 

Additional direct experimental evidence for the Mott-Hubbard state
could be obtained by a careful study of adatom ($3p_z$, $3p_z$)
Auger spectral intensities, which should be easily singled out owing
to the large gaps.
The probability of double occupancy of the adatom dangling bond orbital 
should be almost completely suppressed, dropping from the band value 
of $1/4$, to a value of order $t/U$ 
-- where $t$ is the hopping matrix element and $U$ is the
effective on-site Coulomb repulsion -- which is one order of magnitude smaller. 
Even considering that only half of the orbital
is adatom $3p_z$, this surface should show 
a ($3p_z$, $3p_z$) Auger intensity which, by comparison
with the remaining $(3p,3p)$ values, is anomalously small, 
as a proof of its Mott-Hubbard state.

Summarizing, we have argued that the Mott-Hubbard insulating surface state
in SiC(0001) should leave clear and detectable intra-atomic fingerprints. 
The Si $2p$ core hole will exhibit a $6$-fold multiplet resulting from
the joint effect of intra-atomic exchange, asymmetry, and spin-orbit, 
with a characteristic three-peaked structure. 
Auger intensities should also be affected. 
More experimental and theoretical effort is
clearly called for to check these strong correlations and 
related magnetic effects, 
possibly the first of this magnitude to be suggested 
for an sp-bonded, valence semiconductor surface. 

We thank S. Modesti for useful discussions.
Work at SISSA was partially supported by INFM through PRA LOTUS and HTSC,
by MURST through COFIN97, 
and by the EU, through contract FULPROP ERBFMRXCT970155.


\begin{center} {\bf Figures} \end{center}

\end{document}